# Atom-selective spin-polarized transport in a charge-ordered altermagnet

Liu Yang,[1, 2] Yuan-Yuan Jiang,[3] Xiao-Yan Guo,[1,*] Yi-Dong Liu,[1, 2] Xian-Zhe Chen,[4] Wen-Jian Lu,[1] Yu-Ping Sun,[5,1,6] Ming Li,[7,†] and Ding-Fu Shao[1,§]

[1] *Key Laboratory of Materials Physics, Institute of Solid State Physics, HFIPS, Chinese Academy of Sciences, Hefei 230031, China*

[2] *University of Science and Technology of China, Hefei 230026, China*

[3] *Anhui Provincial Key Laboratory of Magnetic Functional Materials and Devices, School of Materials Science and Engineering, Anhui University, Hefei 230601, China*

[4] *Frontier Institute of Chip and System, State Key Laboratory of Integrated Chips and Systems, Zhangjiang, Fudan International Innovation Center, Fudan University, Shanghai 200433, China*

[5] *Anhui Key Laboratory of Low-Energy Quantum Materials and Devices, High Magnetic Field Laboratory, HFIPS, Chinese Academy of Sciences, Hefei 230031, China*

[6] *Collaborative Innovation Center of Microstructures, Nanjing University, Nanjing 210093, China*

[7] *University of Electronic Science and Technology of China, Chengdu 611731, China*

[†] xyguo@issp.ac.cn; [‡] ming.li7@uestc.edu.cn; [§] dfshao@issp.ac.cn

Altermagnets provide a promising platform for spin-polarized transport without net magnetization, but their transport properties are usually discussed in terms of momentum-space spin splitting. Here, using first-principles calculations and quantum transport simulations, we show that the charge-ordered altermagnet α-$Fe_2PO_5$ exhibits a distinct form of real-space spin selectivity despite weak altermagnetic spin splitting near the Fermi level. The charge order creates inequivalent $Fe^{2+}$ and $Fe^{3+}$ sites within each sublattice, while the puckered C-type antiferromagnetic stacking suppresses inter-sublattice transport. As a result, electron and hole doping activate spin-polarized transport predominantly through $Fe^{3+}$- and $Fe^{2+}$-based channels, respectively. These atom-selective channels carry opposite spin polarizations on the two antiferromagnetic sublattices, giving rise to a globally compensated charge current with hidden Néel spin character. We further propose an all-in-one α-$Fe_2PO_5$ tunnel junction, where matching or mismatching atom-selective conduction channels yields orders-of-magnitude conductance modulation. Our findings establish a real-space design principle for atomically controlled spin functionality and spintronic devices.

The study of charge and spin transport in magnetic materials has long been dominated by a momentum-space perspective, where electronic band structures provide a powerful framework for understanding macroscopic conduction properties. In particular, spin-polarized transport in ferromagnets is well explained by spin-split bands, leading to phenomena such as magnetoresistance [1-8] and spin-transfer torques. [9-11] This approach has been successfully extended to antiferromagnetic (AFM) systems in recent years, where symmetry analysis has revealed unconventional AFM phases—such as non-collinear antiferromagnets [12-19] and altermagnets [20-31]—that exhibit momentum-dependent spin splitting even in the absence of net magnetization. These discoveries have established that antiferromagnets can also support spin-dependent transport properties, enabling high-performance spintronic devices such as AFM tunnel junctions. [14-18,32-35]

While such a momentum-space description effectively captures macroscopic transport signatures, it often over-looks the local and atomically-resolved character of elec-tron flow. In real materials, spin-polarized currents may propagate along specific atomic paths, guided by the local magnetic order, the stacking sequence of magnetic layers, and the valence states and orbital occupations of constituent atoms. These microscopic details are averaged out in con-ventional band-theory approaches, yet they are crucial for understanding spatially heterogeneous transport, especially in doped or chemically complex systems where electronic states become highly localized.

Importantly, this real-space perspective has already yielded significant insights into the nature of local spin transport. [36-40] For instance, in collinear antiferromagnets with specific magnetic stackings—such as the C-type, A-type, and X-type orders—the arrangement of magnetic sublattices directly governs the emergence and pathway of sublattice-confined spin-polarized currents, i.e., the Néel spin currents. [36] These phenomena underscore that the real-space magnetic architecture is a fundamental determinant of spin current propagation, providing a critical complement to the momentum-space picture.

To further explore emergent local transport phenomena in complex magnetic structures, we investigate α-$Fe_2PO_5$—a collinear altermagnetic insulator with a puckered C-type stacking and a charge-ordered state of $Fe^{2+}$ and $Fe^{3+}$ cations that distorts the $FeO_6$ octahedral chains. [41,42] Based on first-principles calculations, we demonstrate that this system hosts doping-dependent, atom-selective Néel spin currents. Although the structural distortion induces strong in-plane anisotropy and associated d-wave altermagnetism, the spin splitting near the Fermi level remains weak due to suppressed intra-sublattice coupling. Remarkably, calculations of local conductance reveal that the spin current flows exclusively along distinct atomic channels depending on doping type: predominantly through $Fe^{3+}$

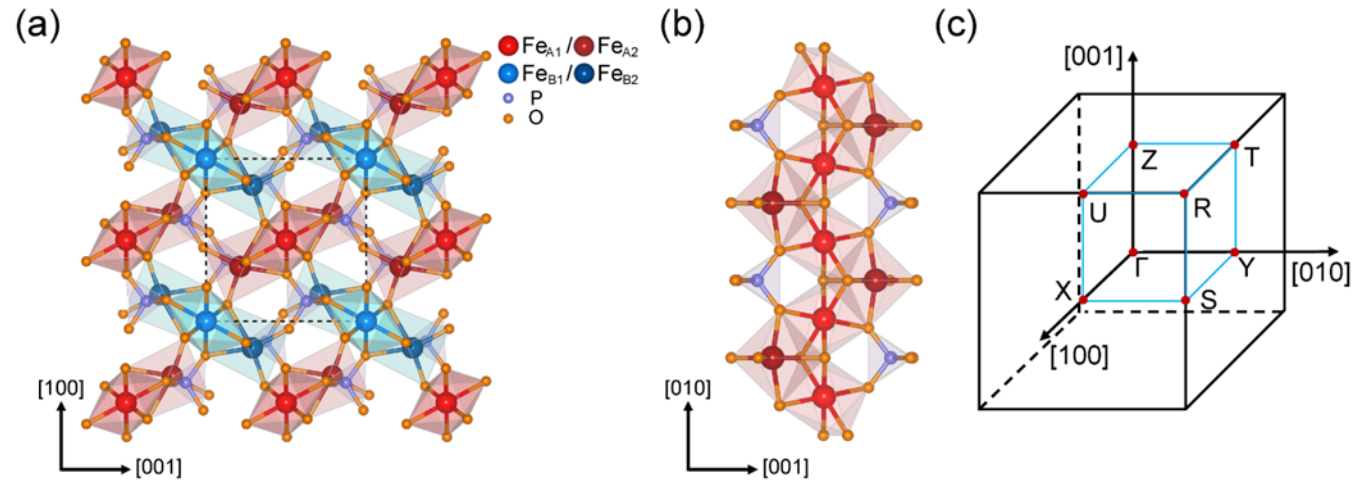

**Fig. 1:** (a,b) The crystal structures of α-$Fe_2PO_5$ with the coordination polyhedra around Fe emphasized in (a) top view and (b) side view of $Fe_AO_6$ octahedra. The dashed black lines indicate the unit cell. (c) The first Brillouin zone of α-$Fe_2PO_5$.

-based channels under electron doping and $Fe^{2+}$-based channels under hole doping. This behavior cannot be inferred from band dispersions or momentum-space spin splitting alone; it requires resolving how site-selective electronic states are connected into continuous transport pathways in real space. By combining real-space magnetic structure analysis with atom-resolved electronic states, we elucidate how the local stacking environment and atomic valence govern the spin-current pathways. Building on this atom-selective transport mechanism, we further propose a proof-of-principle all-in-one tunnel junction based on α-$Fe_2PO_5$ and show that orders-of-magnitude conductance modulation can be achieved by controlling the matching and mismatching of real-space atomic conduction channels. Our findings not only deepen the understanding of microscopically heterogeneous spin transport, but also provide guidelines for designing materials with atomically controlled spin conduction for functional spintronic devices.

First-principles calculations were performed within the framework of density functional theory (DFT) [43] using the QuantumATK. [44] The calculations employed a spin-polarized GGA+U functional [45,46] with an effective Hubbard U = 3 eV applied to the Fe 3*d* orbitals, non-relativistic Fritz-Haber Institute (FHI) pseudopotentials, [47,48] and a plane-wave cutoff energy of 80 Ry. Quantum transport calculations were performed using the non-equilibrium Green's function (NEGF) formalism. [49,50] A k-point grid of 11×11×101 was used along the transport direction.

α-$Fe_2PO_5$ crystallizes in an orthorhombic structure with space group *Pnma* and exhibits AFM order below 250 K. Its crystal and magnetic structures are shown in Fig. 1. The projection onto the (010) plane (Fig. 1a) highlights the pronounced in-plane structural anisotropy and reveals two magnetically inequivalent sublattices, denoted as $Fe_A$ and $Fe_B$ within the unit cell. As illustrated in the side view in Fig. 1(b), each magnetic sublattice contains two crystallographically distinct Fe sites with different valence states. Specifically, the $Fe_{X1}$ (X = A, B) sites are occupied by $Fe^{2+}$ ions, whose $FeO_6$ octahedra share O-O edges to form one-dimensional chains extending along the [010] axis. The $Fe_{X2}$ sites host $Fe^{3+}$ ions with $FeO_6$ octahedra sharing faces with the $Fe^{2+}$ chains on one side and alternate with $PO_4$ tetrahedra on the other, collectively constructing a three-dimensional framework. Correspondingly, the calculated magnitudes of magnetic moments for $Fe_{X1}$ and $Fe_{X2}$ are 3.89 μB and 4.23 μB, respectively, in good agreement with experimental data. [41,42]

α-$Fe_2PO_5$ adopts a puckered *C*-type AFM configuration. Magnetic moments are ferromagnetically aligned along each puckered chain running along [010], while neighboring chains belonging to different magnetic sublattices are coupled antiferromagnetically. The Néel vector lies along the chain direction. Within a given sublattice, $Fe^{2+}O_6$ and $Fe^{3+}O_6$ octahedra are connected via face-sharing, producing strong intra-sublattice magnetic coupling, whereas octahedra belonging to different magnetic sublattices are linked only through corner-sharing oxygen atoms, resulting in much weaker inter-sublattice coupling.

The spin space group of α-$Fe_2PO_5$ is $P^{-1}n^{1}m^{-1}a^{\infty_{010}}m^{1}$, in which the two magnetic sublattices break the combined space-time inversion symmetry but are related by a rotation-translation symmetry operation, placing α-$Fe_2PO_5$ in the class of altermagnets and allowing spin-split band structure. This symmetry allows momentum-dependent spin splitting while simultaneously imposing strong constraints on its magnitude and distribution in reciprocal space, as illustrated in Figs. 2 and 3.

Fig. 2 shows the calculated band structure and density of states (DOS) near the Fermi level ($E_F$) for α-$Fe_2PO_5$. The system exhibits an overall semiconducting state, with bands remaining nearly spin-degenerate along most high-symmetry directions over the Brillouin zone (Fig. 1c). Notably, along the Γ-U direction, finite but small spin splitting appears in both the conduction and valence bands, as highlighted in Fig. 2(b). Near the conduction-band minimum (CBM) and valence-band maximum (VBM), the spin splitting reaches approximately 6.9 and 9.8 meV.

To further characterize the weak altermagnetic splitting in momentum space, we examine the spin-resolved constant-energy surfaces and the corresponding conducting channels at four representative energies, $E_n^1$ and $E_n^2$ in the conduction band and $E_p^1$ and $E_p^2$ in the valence band, as marked by the blue dashed lines in Fig. 2(c). As shown in the top two rows of Fig. 3, the spin-up and spin-down constant-energy surfaces have nearly identical shapes at all selected energies, consistent with the small spin splitting observed in the band structure in Fig. 2(b).

We further project these states onto the two-dimensional transverse Brillouin zone perpendicular to [010] transport direction to calculate the transverse-momentum-resolved spin polarization, $p_{\parallel}(\mathbf{k}_{\parallel}) = \frac{N_{\parallel}^{\uparrow}-N_{\parallel}^{\downarrow}}{N_{\parallel}^{\uparrow}+N_{\parallel}^{\downarrow}}$, where $N_{\parallel}^{\uparrow}$ and $N_{\parallel}^{\downarrow}$ are the transverse-momentum-resolved numbers of conducting channels. As shown in the bottom row of Fig. 3, the nonzero spin polarization occupies only a small portion of the transverse

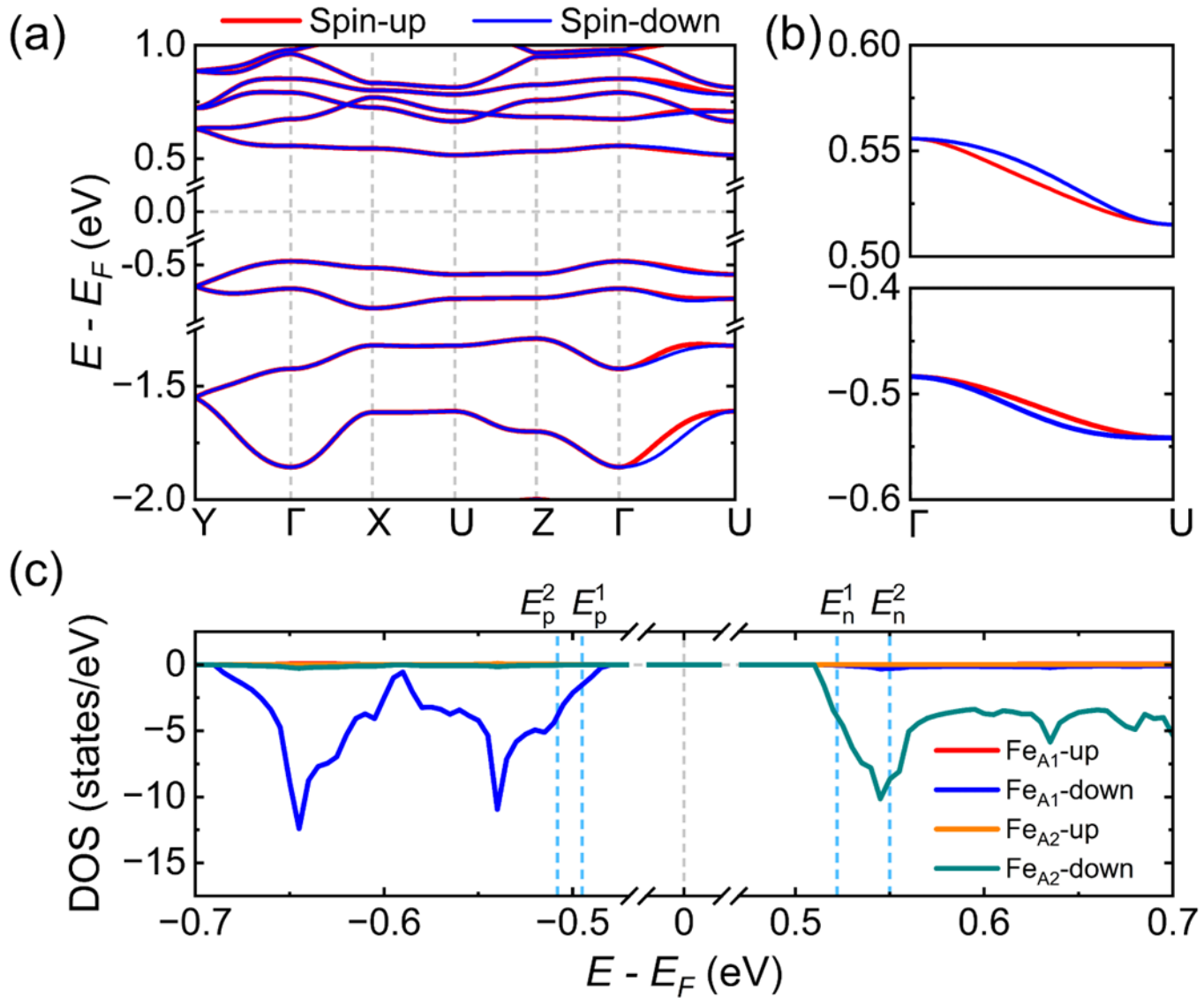

**Fig. 2:** (a) The band structure of α-$Fe_2PO_5$. (b) The en-larged band structure along the Γ-U high-symmetry path. (c) The $Fe_A$-resolved density of states of contributed by the valence and conduction bands.

Brillouin zone, confirming that the altermagnetic spin splitting is very weak.

It is commonly believed that strong surrounding environments of magnetic atoms favor stronger altermagnetic spin splitting. In this context, the exceptionally weak spin splitting observed in α-$Fe_2PO_5$ is surprising, especially given its highly anisotropic sublattice structure. In fact, a simple lattice model clarifies that strong *d*-wave spin splitting depends crucially on anisotropic hopping between chains within the same sublattice in the plane. [37] As shown in the top view of α-$Fe_2PO_5$ (Fig. 1a), its puckered C-type stacking leads to a layout where chains of the same sublattice are separated from each other by chains of the other sublattice. This results in extremely poor in-plane connectivity within each sublattice. Consequently, hopping between chains of the same sublattice is highly suppressed in all in-plane directions, and the negligible variation among these weak hopping pathways leads to the remarkably small spin splitting.

Despite the weak spin splitting in momentum space, we find that α-$Fe_2PO_5$ exhibits exotic spin-dependent transport in real space, due to its puckered C-type stacking in conjunction with the local electronic structure of Fe atoms. Fig. 2(c) presents the atom- and spin-resolved density of states (DOS) for the two inequivalent Fe sites within the $Fe_A$ sublattice. The conduction bands are dominated by spin-down $Fe^{3+}$ 3*d* orbitals, whereas the valence-band states are mainly contributed by spin-down $Fe^{2+}$ 3*d* orbitals. Notably, the spin character of the DOS near the Fermi level ($E_F$) is opposite to the direction of the local magnetic moments, which can be qualitatively understood within a double-exchange picture. In the undistorted mixed-valence state where equivalent Fe atoms host a mixed-valence of +2.5, an itinerant *d* electron shared between the Fe sites must align antiparallel to the five localized *d* electrons. Then, the structural distortion lifts this equivalence, splitting the bands and producing $Fe_{X1}$ and $Fe_{X2}$ sites, yet the spin character of the states at both the CBM and VBM remains antiparallel to the local moments.

Therefore, although the total DOS is spin-degenerate and the spin splitting in the band structure is very weak, the electronic states near the band edges exhibit strong site- and spin-selective features. Consequently, electrons doped into the CBM or holes doped into the VBM predominantly occupy a single spin channel localized on a specific Fe valence site, providing the microscopic prerequisite for sublattice-dependent spin transport and facilitating the emergence of Néel-type spin currents in α-$Fe_2PO_5$.

To explicitly demonstrate how charge or spin currents are distributed among inequivalent atomic sites in real space, we hereby turn to an analysis of the sublattice-resolved transport by computing the bond-resolved charge transmission $T_{ij} = T_{ij}^{\uparrow} + T_{ij}^{\downarrow}$ (Figs. 4a and b) and the corresponding spin-resolved bond transmission $T_{ij}^{S} = T_{ij}^{\uparrow} - T_{ij}^{\downarrow}$ (Figs. 4c and d) along the [010] direction, following Ref. [37]. We use $E_n^2$ ( $E_p^2$ ) as the representative energy to show the case of the *n*-doped (*p*-doped) α-$Fe_2PO_5$. For $E_n^2$, charge transmission is dominated by hopping within ferromagnetically aligned chains formed by $Fe^{3+}$ ions (Fig. 4a). In contrast, for $E_p^2$, transport proceeds primarily along $Fe^{2+}$ chains. This valence-dependent transport behavior directly reflects the orbital-resolved DOS in Fig. 2(c), where CBM (VBM) states originate predominantly from $Fe^{3+}$ ($Fe^{2+}$) ions. In both doping regimes, charge transmission between neighboring $Fe_A$ and $Fe_B$ chains is strongly suppressed. This suppression arises from the large inter-chain separation and the corner-sharing connectivity between sublattices, which significantly weakens inter-sublattice hopping. Consequently, charge transport in α-$Fe_2PO_5$ is effectively confined within individual magnetic sublattices and proceeds predominantly along the chain direction.

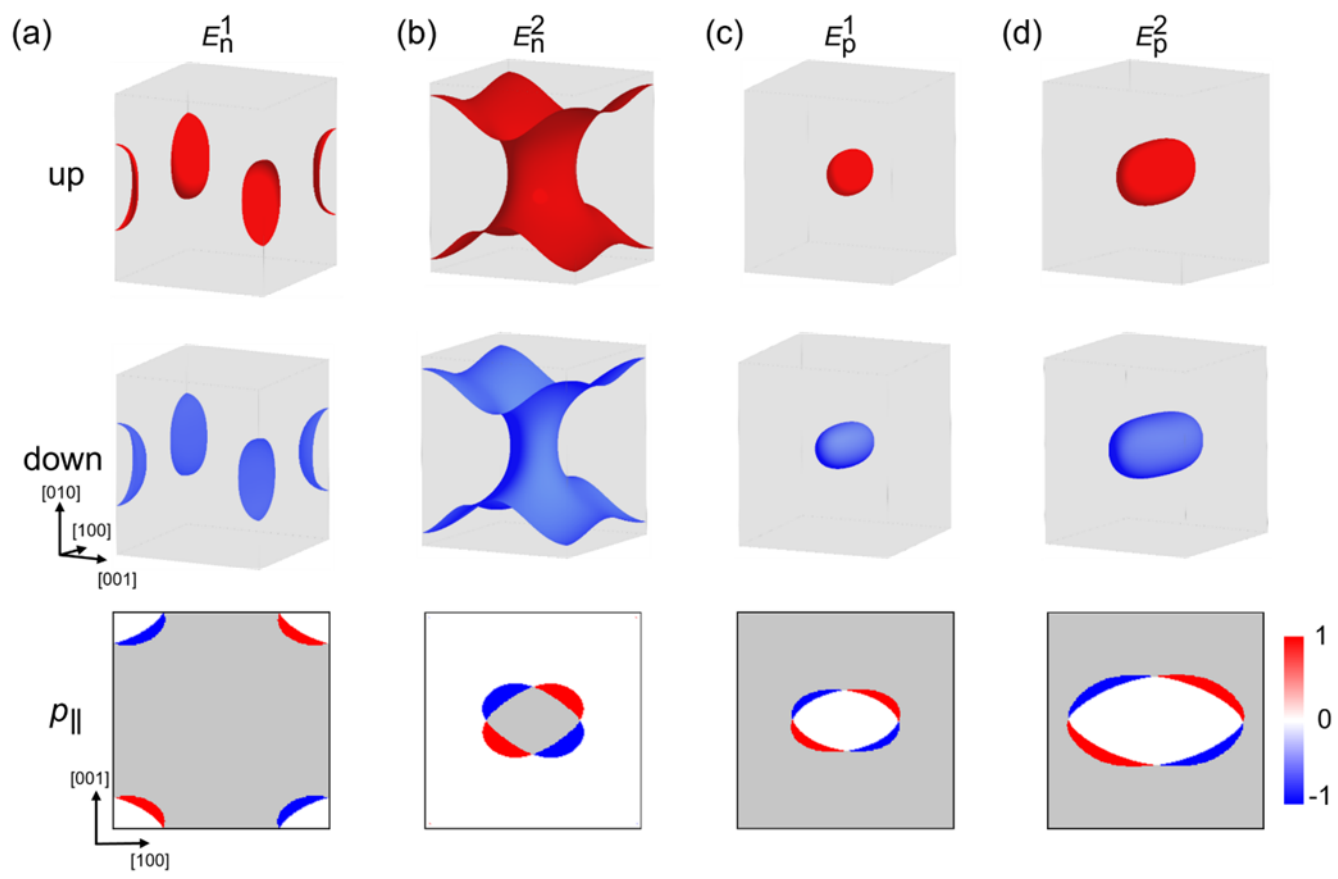

**Fig. 3:** (a-d) Three-dimensional constant-energy surfaces for spin-up and spin-down states at representative *n*- and *p*-doped energies and the $\mathbf{k}_{\parallel}$-resolved spin polarization $p_{\parallel}(\mathbf{k}_{\parallel})$ for transport along the [010] direction. Gray regions denote momenta without available conducting channels, where $N_{\parallel}^{\uparrow} = N_{\parallel}^{\downarrow} = 0$.

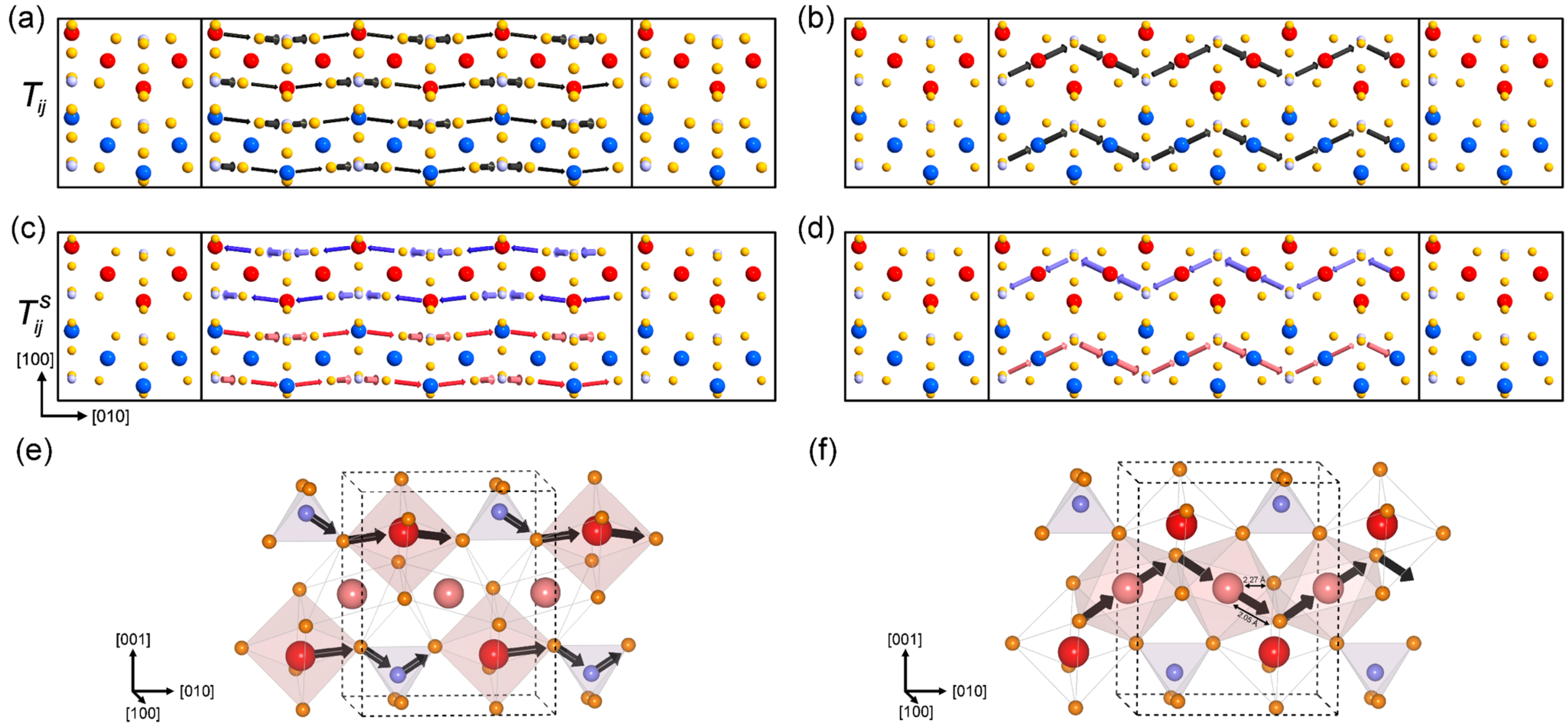

**Fig. 4:** Real-space bond-resolved transport in doped α-$Fe_2PO_5$. (a,b) Bond charge transmission $T_{ij}$ and (c,d) bond spin transmission $T^s_{ij}$ computed at the representative *n*- and *p*-doped energies $E^2_n$ and $E^2_p$, respectively, as marked by the blue dashed lines in Fig. 2(c). Panels (a,c) correspond to electron doping, while (b,d) correspond to hole doping. The arrow radius represents the magnitude of $T_{ij}$ or $T^s_{ij}$; in (c,d), and the color indicates the sign of spin transmission. (e,f) Enlarged views of the dominant transmission pathways under electron doping (e) and hole doping (f) within the unit cell. Octahedra not involved in the primary transmission are shown as transparent.

The corresponding spin-resolved bond transmission (Figs. 4c and 4d) further reveals pronounced sublattice and valence selectivity. Spin transport is almost entirely confined to intra-sublattice bonds, demonstrating that spin-polarized carriers propagate coherently along the ferromagnetically aligned chains. Electron doping activates spin transport predominantly on $Fe^{3+}$ sites, whereas hole doping restricts it to $Fe^{2+}$ sites, which, again, reflects the site-resolved spin polarization of the electronic states (Fig. 2c). This spin selectivity of the electronic states near $E_F$ effectively acts as a spin-dependent filter that blocks spin-conserving hopping between sublattices, which strongly suppresses the inter-sublattice spin transport, even though $Fe_A$ and $Fe_B$ chains are structurally connected through shared oxygen atoms. Consequently, α-$Fe_2PO_5$ supports highly spin-polarized and valence-ordered spin transport along the chain direction. An electric field applied along [010] drives parallel charge currents in $Fe_A$ and $Fe_B$ chains but with opposite spin polarizations, giving rise to robust Néel spin currents within the globally spin neutral charge current.

Visualization of the real-space current distribution within the unit cell (Figs. 4e and 4f) further clarifies the microscopic hopping mechanisms. Under electron doping, electrons hop between neighboring $Fe^{3+}O_6$ octahedra via intervening $PO_4$ tetrahedra, forming the shortest available hopping path. In contrast, hole-doped transport is dominated by direct hopping between adjacent $Fe^{2+}O_6$ octahedra through shared O-O edges. These distinct hopping geometries underpin the observed atom-selective spin-current channels.

This doping-tunable, atom-selective Néel spin current presents intriguing opportunities for novel physics and device applications. As a proof of principle, we consider an all-in-one α-$Fe_2PO_5$ tunnel junction along the [010] direction, in which the left and right regions are charge doped and serve as conducting electrodes, while the central pristine region remains insulating and acts as the tunnel barrier, as schematically shown in Fig. 5(a). The doping type on each side may be controlled independently, for example, by electrostatic or ionic-liquid gating, allowing the active atomic conduction channel to be selected separately in the two electrodes. Electron doping activates $Fe^{3+}$-based channels, whereas hole doping activates $Fe^{2+}$-based channels. Therefore, identical doping on the two sides produces matched real-space channels, while opposite doping produces mismatched channels.

We consider two representative doping regimes: $E^1_n$ and $E^1_p$ qualitatively represent shallow *n*- and *p*- doping, whereas $E^2_n$

**Table 1**: The total transmissions ($T$) for the all-in-one α-$Fe_2PO_5$ tunnel junction.

| | $n_1n_1$ | $p_1p_1$ | $n_1p_1$ | $n_2n_2$ | $p_2p_2$ | $n_2p_2$ |
|---|---|---|---|---|---|---|
| $T$ | $9.9\times10^{-3}$ | $9.9\times10^{-5}$ | $2.5\times10^{-12}$ | $1.4\times10^{-2}$ | $6.3\times10^{-4}$ | $2.0\times10^{-5}$ |

and $E_p^2$ represent a deeper-doping regime. A common feature in both regimes is that the transmission is much larger when the two electrodes have the same doping type, namely $n_1n_1$, $p_1p_1$, $n_2n_2$, and $p_2p_2$, than when they have opposite doping types, namely $n_1p_1$ and $n_2p_2$, as listed in Table 1. Thus, switching the relative doping polarity of the two electrodes provides a direct way to switch the junction between high- and low-transmission states

If one only considers the shallow-doping case, the suppressed transmission in the $n_1p_1$configuration may appear to originate from a momentum-space channel mismatch. As shown in Fig. 5(b,c), at the $n$-doped energies $E_n^1$, the conducting channels appear away from the zone center, while at the hole-doped energies $E_p^1$, the available conducting channels are mainly located around the Brillouin-zone center. Consequently, the left and right electrodes in the $n_1p_1$ configuration have the negligible overlap in their available $\mathbf{k}_\parallel$-resolved conducting channels, naturally leading to weak transmission (Fig. 5d).

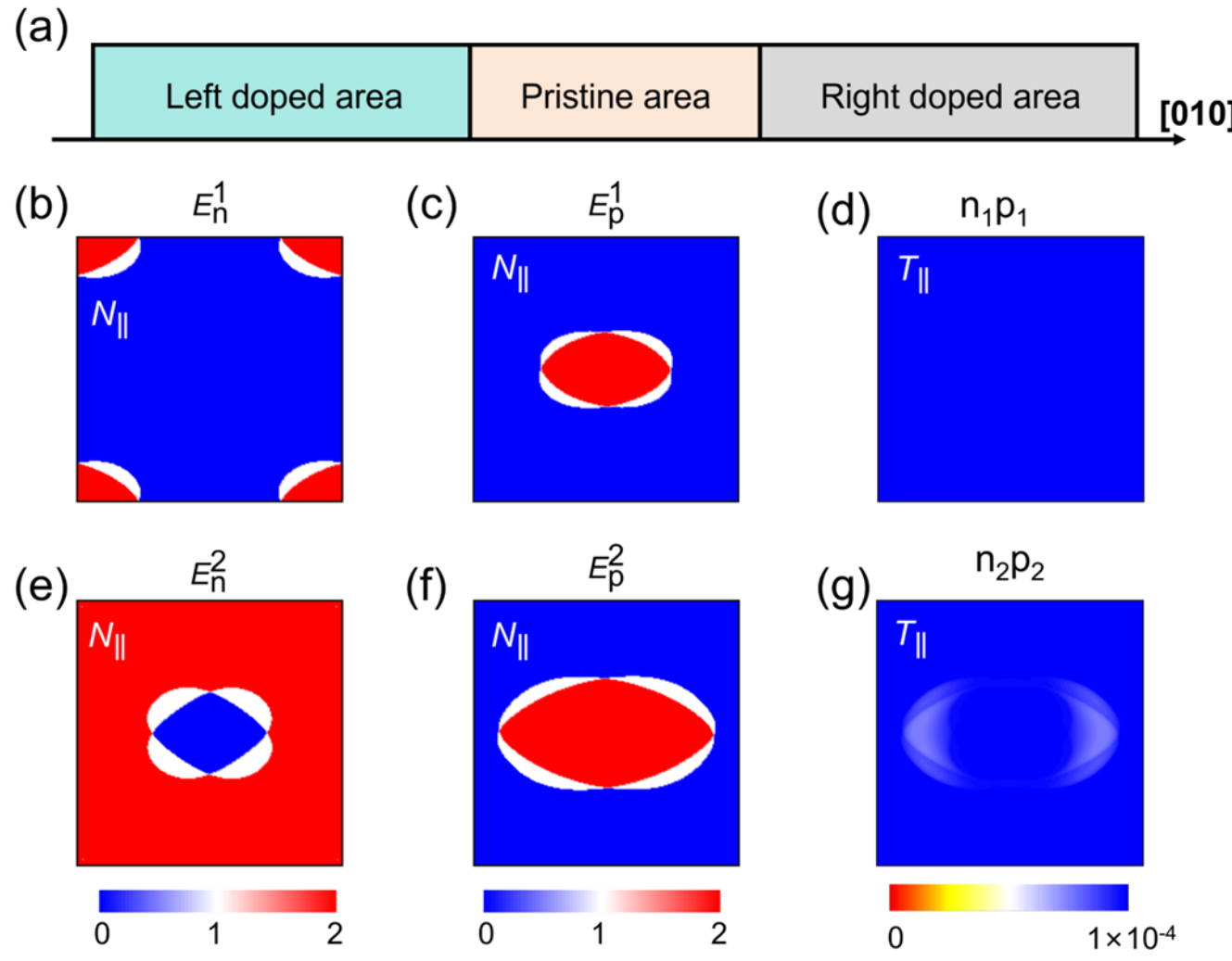

**Fig. 5:** (a) Schematic of the proposed α-$Fe_2PO_5$ junction, consisting of left and right doped regions separated by a pristine insulating region. The labels $n_1n_1$, $n_2n_2$, $p_1p_1$, and $p_2p_2$ denote junctions in which both electrodes are doped to the same representative electron- or hole-doped energies, $E_n^1$, $E_n^2$, $E_p^1$, and $E_p^2$, respectively. The labels $n_1p_1$ and $n_2p_2$ denote junctions in which the left electrode is electron-doped to $E_n^1$ or $E_n^2$, while the right electrode is hole-doped to $E_p^1$ or $E_p^2$, respectively. (b,c) $\mathbf{k}_\parallel$-resolved numbers of total conduction channels $N_\parallel = N_\parallel^\uparrow + N_\parallel^\downarrow$ at $E_n^1$ (b) and $E_p^1$ (c) for transport along the [010] direction. (d) Transverse-momentum-resolved transmission $T_\parallel$ for the doping configuration $n_1p_1$. (e,f) $N_\parallel$ at $E_n^2$ (e) and $E_p^2$ (f) for transport along the [010] direction. (g) $T_\parallel$ for the doping configuration $n_2p_2$.

However, this momentum-space picture alone cannot explain the deeper-doping case. While the transmission in the $n_2p_2$ configuration is indeed larger than that of $n_1p_1$ due to the deeper doping level, it remains significantly lower compared to the case of $n_2n_2$ or $p_2p_2$ (Table 1). In the $n_2p_2$ configuration, the $n$- and $p$-doped electrodes exhibit a partial overlap of their $\mathbf{k}_\parallel$-resolved conducting channels (Figs. 5e and f). Nevertheless, the transmission remains strongly suppressed even within the overlapping region. This behavior demonstrates that the ON/OFF effect is not governed solely by momentum-space channel overlap. Instead, its more fundamental origin lies in the mismatch of real-space atom-selective transport pathways. Electron doping activates $Fe^{3+}$-based channels, whereas hole doping activates $Fe^{2+}$-based channels. Because these two types of atomic channels are spatially distinct and only weakly connected, a carrier injected from an $Fe^{3+}$-dominated channel on one side cannot efficiently continue into an $Fe^{2+}$-dominated channel on the other side. Therefore, matched doping preserves continuous real-space atomic pathways and yields a high-transmission state, whereas opposite doping breaks this channel continuity and produces a low-transmission state.

These results establish real-space atom-selective channel matching as the central mechanism behind the large ON/OFF modulation in the proposed α-$Fe_2PO_5$ junction (Table 1). Although relatively large doping levels are used here to make the contrast between the two channel types explicit, the essential channel selectivity already appears near the band edges, suggesting that sizeable conductance modulation may be achievable at lower carrier densities.

Moreover, since doping renders the DOS of either $Fe^{2+}$ or $Fe^{3+}$ atoms fully spin-polarized, an injected spin current can flow only into one type of spin-polarized atomic channel. This enables spin torque to be precisely applied to specific atoms within a chosen sublattice. Such a highly asymmetric spin torque is expected to induce unconventional magnetic dynamics.[51] Experimentally, such atom-selective Néel spin currents can be detected in real space using spin-polarized scanning tunneling microscopy (SP-STM). [52]

In summary, we have demonstrated atom-selective spin-polarized transport in the charge-ordered altermagnet α-$Fe_2PO_5$. Although the altermagnetic spin splitting near the Fermi level is weak, charge-order-induced site-selective band-edge states and suppressed inter-sublattice connectivity enable electron and hole doping to activate distinct $Fe^{3+}$- and $Fe^{2+}$-based transport channels. The opposite spin polarizations on the two antiferromagnetic sublattices give the compensated charge current a hidden Néel spin character. The proposed all-in-one junction further shows that real-space channel matching can be converted into a large conductance modulation. These findings not only highlight the decisive role of real-space magnetic connectivity and valence order in spin transport of altermagnets but also provide a design principle for spintronics.

**Acknowledgments.** This work was supported by the National Key R&D Program of China (Grant No. 2024YFB3614100 and 2022YFA1403200), the National Natural Science Foundation of China (Grants Nos. 12241405, 12274411, and 12504098), the Basic Research Program of the Chinese Academy of Sciences Based on Major Scientific Infrastructures (Grant No. JZHKYPT-2021-08), and the CAS Project for Young Scientists in Basic Research (Grant No. YSBR-084). The calculations were performed at Hefei Advanced Computing Center.

[1] M. N. Baibich, J. M. Broto, A. Fert, F. N. Van Dau, F. Petroff, P. Etienne, G. Creuzet, A. Friederich, and J. Chazelas, Giant Magnetoresistance of (001)Fe/(001)Cr Magnetic Superlattices, Phys. Rev. Lett. **61**, 2472 (1988).

[2] J. S. Moodera, L. R. Kinder, T. M. Wong, and R. Meservey, Large Magnetoresistance at Room Temperature in Ferromagnetic Thin Film Tunnel Junctions, Phys. Rev. Lett. **74**, 3273 (1995).

[3] S. Yuasa, T. Nagahama, A. Fukushima, Y. Suzuki, and K. Ando, Giant room-temperature magnetoresistance in single-crystal Fe/MgO/Fe magnetic tunnel junctions, Nat. Mater. **3**, 868 (2004).

[4] M. Younis, M. Abdullah, S. Dai, M. A. Iqbal, W. Tang, M. T. Sohail, S. Atiq, H. Chang, and Y.-J. Zeng, Magnetoresistance in 2D Magnetic Materials: From Fundamentals to Applications, Adv. Funct. Mater. **35**, 2417282 (2025).

[5] M. Julliere, Tunneling between ferromagnetic films, Phys. Lett. A **54**, 225 (1975).

[6] Y. T. Evgeny, N. M. Oleg, and R. L. Patrick, Spin-dependent tunnelling in magnetic tunnel junctions, J. Phys. Condens. Matter **15**, R109 (2003).

[7] P. Mavropoulos, N. Papanikolaou, and P. H. Dederichs, Complex Band Structure and Tunneling through Ferromagnet/Insulator/Ferromagnet Junctions, Phys. Rev. Lett. **85**, 1088 (2000).

[8] S. S. P. Parkin, C. Kaiser, A. Panchula, P. M. Rice, B. Hughes, M. Samant, and S.-H. Yang, Giant tunnelling magnetoresistance at room temperature with MgO (100) tunnel barriers, Nat. Mater. **3**, 862 (2004).

[9] J. C. Slonczewski, Current-driven excitation of magnetic multilayers, J. Magn. Magn. Mater. **159**, L1 (1996).

[10] A. Brataas, A. D. Kent, and H. Ohno, Current-induced torques in magnetic materials, Nat. Mater. **11**, 372 (2012).

[11] D. C. Ralph and M. D. Stiles, Spin transfer torques, J. Magn. Magn. Mater. **320**, 1190 (2008).

[12] J. Železný, Y. Zhang, C. Felser, and B. Yan, Spin-Polarized Current in Noncollinear Antiferromagnets, Phys. Rev. Lett. **119**, 187204 (2017).

[13] G. Gurung, D.-F. Shao, and E. Y. Tsymbal, Transport spin polarization of noncollinear antiferromagnetic antiperovskites, Phys. Rev. Mater. **5**, 124411 (2021).

[14] J. Dong, X. Li, G. Gurung, M. Zhu, P. Zhang, F. Zheng, E. Y. Tsymbal, and J. Zhang, Tunneling Magnetoresistance in Noncollinear Antiferromagnetic Tunnel Junctions, Phys. Rev. Lett. **128**, 197201 (2022).

[15] P. Qin, H. Yan, X. Wang, H. Chen, Z. Meng, J. Dong, M. Zhu, J. Cai, Z. Feng, X. Zhou, L. Liu, T. Zhang, Z. Zeng, J. Zhang, C. Jiang, and Z. Liu, Room-temperature magnetoresistance in an all-antiferromagnetic tunnel junction, Nature **613**, 485 (2023).

[16] X. Chen, T. Higo, K. Tanaka, T. Nomoto, H. Tsai, H. Idzuchi, M. Shiga, S. Sakamoto, R. Ando, H. Kosaki, T. Matsuo, D. Nishio-Hamane, R. Arita, S. Miwa, and S. Nakatsuji, Octupole-driven magnetoresistance in an antiferromagnetic tunnel junction, Nature **613**, 490 (2023).

[17] G. Gurung, M. Elekhtiar, Q.-Q. Luo, D.-F. Shao, and E. Y. Tsymbal, Nearly perfect spin polarization of noncollinear antiferromagnets, Nat. Commun. **15**, 10242 (2024).

[18] Q.-Q. Luo, X.-Y. Guo, H. Zhou, G. Gurung, J.-M. Xu, W.-J. Lu, Y.-P. Sun, E. Y. Tsymbal, and D.-F. Shao, Angular-dependent tunneling magnetoresistance in a tunnel junction with ferromagnetic and noncollinear antiferromagnetic electrodes, Phys. Rev. B **111**, 144417 (2025).

[19] B. H. Rimmler, B. Pal, and S. S. P. Parkin, Non-collinear antiferromagnetic spintronics, Nat. Rev. Mater. **10**, 109 (2025).

[20] C. Wu, K. Sun, E. Fradkin, and S.-C. Zhang, Fermi liquid instabilities in the spin channel, Phys. Rev. B **75**, 115103 (2007).

[21] R. González-Hernández, L. Šmejkal, K. Výborný, Y. Yahagi, J. Sinova, T. Jungwirth, and J. Železný, Efficient Electrical Spin Splitter Based on Nonrelativistic Collinear Antiferromagnetism, Phys. Rev. Lett. **126**, 127701 (2021).

[22] S. Hayami, Y. Yanagi, and H. Kusunose, Momentum-Dependent Spin Splitting by Collinear Antiferromagnetic Ordering, J. Phys. Soc. Jpn. **88**, 123702 (2019).

[23] L.-D. Yuan, Z. Wang, J.-W. Luo, E. I. Rashba, and A. Zunger, Giant momentum-dependent spin splitting in centrosymmetric low-*Z* antiferromagnets, Phys. Rev. B **102**, 014422 (2020).

[24] L.-D. Yuan, Z. Wang, J.-W. Luo, and A. Zunger, Prediction of low-Z collinear and noncollinear antiferromagnetic compounds having momentum-dependent spin splitting even without spin-orbit coupling, Phys. Rev. Mater. **5**, 014409 (2021).

[25] S. Hayami, Y. Yanagi, and H. Kusunose, Bottom-up design of spin-split and reshaped electronic band structures in antiferromagnets without spin-orbit coupling: Procedure on the basis of augmented multipoles, Phys. Rev. B **102**, 144441 (2020).

[26] P. Liu, J. Li, J. Han, X. Wan, and Q. Liu, Spin-Group Symmetry in Magnetic Materials with Negligible Spin-Orbit Coupling, Phys. Rev. X **12**, 021016 (2022).

[27] L. Šmejkal, J. Sinova, and T. Jungwirth, Beyond Conventional Ferromagnetism and Antiferromagnetism: A Phase with Nonrelativistic Spin and Crystal Rotation Symmetry, Phys. Rev. X **12**, 031042 (2022).

[28] L. Šmejkal, J. Sinova, and T. Jungwirth, Emerging Research Landscape of Altermagnetism, Phys. Rev. X **12**, 040501 (2022).

[29] L.-D. Yuan and A. Zunger, Degeneracy Removal of Spin Bands in Collinear Antiferromagnets with Non-Interconvertible Spin-Structure Motif Pair, Adv. Mater. **35**, 2211966 (2023).

[30] C. Song, H. Bai, Z. Zhou, L. Han, H. Reichlova, J. H. Dil, J. Liu, X. Chen, and F. Pan, Altermagnets as a new class of functional materials, Nat. Rev. Mater. **10**, 473 (2025).

[31] L. Bai, W. Feng, S. Liu, L. Šmejkal, Y. Mokrousov, and Y. Yao, Altermagnetism: Exploring New Frontiers in Magnetism and Spintronics, Adv. Funct. Mater. **34**, 2409327 (2024).

[32] D.-F. Shao, S.-H. Zhang, M. Li, C.-B. Eom, and E. Y. Tsymbal, Spin-neutral currents for spintronics, Nat. Commun. **12**, 7061 (2021).

[33] Y.-Y. Jiang, Z.-A. Wang, K. Samanta, S.-H. Zhang, R.-C. Xiao, W. J. Lu, Y. P. Sun, E. Y. Tsymbal, and D.-F. Shao, Prediction of giant tunneling magnetoresistance in $RuO_2/TiO_2/RuO_2$ (110) antiferromagnetic tunnel junctions, Phys. Rev. B **108**, 174439 (2023).

[34] L. Šmejkal, A. B. Hellenes, R. González-Hernández, J. Sinova, and T. Jungwirth, Giant and Tunneling Magnetoresistance in Unconventional Collinear Antiferromagnets with Nonrelativistic Spin-Momentum Coupling, Phys. Rev. X **12**, 011028 (2022).

[35] K. Samanta, Y.-Y. Jiang, T. R. Paudel, D.-F. Shao, and E. Y. Tsymbal, Tunneling magnetoresistance in magnetic tunnel junctions with a single ferromagnetic electrode, Phys. Rev. B **109**, 174407 (2024).

[36] D.-F. Shao, Y.-Y. Jiang, J. Ding, S.-H. Zhang, Z.-A. Wang, R.-C. Xiao, G. Gurung, W. J. Lu, Y. P. Sun, and E. Y. Tsymbal, Néel Spin Currents in Antiferromagnets, Phys. Rev. Lett. **130**, 216702 (2023).

[37] S.-S. Zhang, Z.-A. Wang, B. Li, Y.-Y. Jiang, S.-H. Zhang, R.-C. Xiao, L.-X. Liu, X. Luo, W.-J. Lu, M. Tian, Y.-P. Sun, E. Y. Tsymbal, H. Du, and D.-F. Shao, X-type stacking in cross-chain antiferromagnets, Newton **1**, 100068 (2025).

[38] Z.-A. Wang, B. Li, S.-S. Zhang, W.-J. Lu, M. Tian, Y.-P. Sun, E. Y. Tsymbal, K. Wang, H. Du, and D.-F. Shao, Giant Uncompensated Magnon Spin Currents in X-Type Magnets, Chin. Phys. Lett. **42**, 110713 (2025).

[39] L. Yang, Y.-Y. Jiang, X.-Y. Guo, S.-H. Zhang, R.-C. Xiao, W.-J. Lu, L. Wang, Y.-P. Sun, E. Y. Tsymbal, and D.-F. Shao, Interface-controlled antiferromagnetic tunnel junctions, Newton **1**, 100142 (2025).

[40] W.-M. Zhao, Y.-L. Liu, L. Yang, C. Tan, Y. Yang, Z. Zhu, M. Chen, T. Yan, R. Hu, J. Partridge, G. Wang, M. Tian, D.-F. Shao, and L. Wang, Interface-controlled antiferromagnetic tunnel junctions based on a metallic van der Waals A-type antiferromagnet, Nat. Commun. **17**, 268 (2025).

[41] A. Modaressi, A. Courtois, R. Gerardin, B. Malaman, and C. Gleitzer, $Fe_2PO_5$, un phosphate de fer de valence mixte. Préparation et études structurale, mössbauer et magnétique, Journal of Solid State Chemistry **40**, 301 (1981).

[42] J. K. Warner, A. K. Cheetham, D. E. Cox, and R. B. Von Dreele, Valence contrast between iron sites in $\alpha$-$Fe_2PO_5$: a comparative study by magnetic neutron and resonant x-ray powder diffraction, J. Am. Chem. Soc. **114**, 6074 (1992).

[43] P. Hohenberg and W. Kohn, Inhomogeneous Electron Gas, Phys. Rev. **136**, B864 (1964).

[44] S. Smidstrup, T. Markussen, P. Vancraeyveld, J. Wellendorff, J. Schneider, T. Gunst, B. Verstichel, D. Stradi, P. A. Khomyakov, U. G. Vej-Hansen, M.-E. Lee, S. T. Chill, F. Rasmussen, G. Penazzi, F. Corsetti, A. Ojanperä, K. Jensen, M. L. N. Palsgaard, U. Martinez, A. Blom, M. Brandbyge, and K. Stokbro, QuantumATK: an integrated platform of electronic and atomic-scale modelling tools, J. Phys. Condens. Matter **32**, 015901 (2020).

[45] J. P. Perdew, K. Burke, and M. Ernzerhof, Generalized Gradient Approximation Made Simple, Phys. Rev. Lett. **77**, 3865 (1996).

[46] S. L. Dudarev, G. A. Botton, S. Y. Savrasov, C. J. Humphreys, and A. P. Sutton, Electron-energy-loss spectra and the structural stability of nickel oxide: An LSDA+U study, Phys. Rev. B **57**, 1505 (1998).

[47] X. Gonze, P. Käckell, and M. Scheffler, Ghost states for separable, norm-conserving, Iab initioP pseudopotentials, Phys. Rev. B **41**, 12264 (1990).

[48] K. Lejaeghere, G. Bihlmayer, T. Björkman, P. Blaha, S. Blügel, V. Blum, D. Caliste, I. E. Castelli, S. J. Clark, A. Dal Corso, S. de Gironcoli, T. Deutsch, J. K. Dewhurst, I. Di Marco, C. Draxl, M. Dułak, O. Eriksson, J. A. Flores-Livas, K. F. Garrity, L. Genovese, P. Giannozzi, M. Giantomassi, S. Goedecker, X. Gonze, O. Grånäs, E. K. U. Gross, A. Gulans, F. Gygi, D. R. Hamann, P. J. Hasnip, N. A. W. Holzwarth, D. Iuşan, D. B. Jochym, F. Jollet, D. Jones, G. Kresse, K. Koepernik, E. Küçükbenli, Y. O. Kvashnin, I. L. M. Locht, S. Lubeck, M. Marsman, N. Marzari, U. Nitzsche, L. Nordström, T. Ozaki, L. Paulatto, C. J. Pickard, W. Poelmans, M. I. J. Probert, K. Refson, M. Richter, G.-M. Rignanese, S. Saha, M. Scheffler, M. Schlipf, K. Schwarz, S. Sharma, F. Tavazza, P. Thunström, A. Tkatchenko, M. Torrent, D. Vanderbilt, M. J. van Setten, V. Van Speybroeck, J. M. Wills, J. R. Yates, G.-X. Zhang, and S. Cottenier, Reproducibility in density functional theory calculations of solids, Science **351**, aad3000 (2016).

[49] J. Taylor, H. Guo, and J. Wang, Ab initio modeling of quantum transport properties of molecular electronic devices, Phys. Rev. B **63**, 245407 (2001).

[50] M. Brandbyge, J.-L. Mozos, P. Ordejón, J. Taylor, and K. Stokbro, Density-functional method for nonequilibrium electron transport, Phys. Rev. B **65**, 165401 (2002).

[51] S.-S. Zhang, Z.-A. Wang, B. Li, W.-J. Lu, M. Tian, Y.-P. Sun, H. Du, and D.-F. Shao, Deterministic Switching of the Néel Vector by Asymmetric Spin Torque, Phys. Rev. Lett. **136**, 096702 (2026).

[52] L. Schneider, P. Beck, J. Wiebe, and R. Wiesendanger, Atomic-scale spin-polarization maps using functionalized superconducting probes, Sci. Adv. **7**, eabd7302 (2021).